\documentclass[sigconf]{acmart}
\AtBeginDocument{%
  }

\setcopyright{none}             
\renewcommand\footnotetextcopyrightpermission[1]{} 

\setcopyright{none}
\renewcommand\footnotetextcopyrightpermission[1]{}

\acmConference[]{}{}{}
\acmBooktitle{}
\AtBeginDocument{%
    \let\origmaketitle\maketitle
    \renewcommand{\maketitle}{%
        \origmaketitle
        \begingroup
        \renewcommand\thefootnote{}\footnote{\noindent \textbf{Author's Copy.} Accepted for publication at the \textit{Proceedings of the ACM Web Conference 2026 (WWW '26)}. This work is licensed under a \textbf{CC BY 4.0}.}%
        \addtocounter{footnote}{-1}%
        \endgroup
    }
}

\usepackage{listings}
\usepackage{xcolor}
\usepackage{graphicx}
\usepackage{subcaption}
\usepackage{listings}
\usepackage{booktabs}
\usepackage{amsmath}
\usepackage{caption}
\usepackage{ragged2e} 
\usepackage{booktabs}
\usepackage{makecell}
\usepackage{threeparttable} 
\usepackage{balance}
\usepackage{lmodern}

\begin{document}

\title{Multilingual Reference Need Assessment System for Wikipedia}

\author{Aitolkyn Baigutanova}
\authornote{This work was conducted while under contract with the Wikimedia Foundation.}
\affiliation{
  \institution{Wikimedia Foundation}
  \city{San Francisco}
  \country{USA}
}
\email{aitolkyn.b@kaist.ac.kr}

\author{Francisco Navas}
\affiliation{
  \institution{Wikimedia Foundation}
  \city{Brooklyn}
  \country{USA}}
\email{fnavas@wikimedia.org}

\author{Pablo Aragón}
\affiliation{
  \institution{Wikimedia Foundation}
  \city{Barcelona}
  \country{Spain}}
\email{paragon@wikimedia.org}

\author{Mykola Trokhymovych}
\affiliation{
  \institution{Pompeu Fabra University}
  \institution{Wikimedia Foundation}
    \city{Barcelona}
  \country{Spain}
}
\email{mykola.trokhymovych@upf.edu}

\author{Muniza Aslam}
\authornotemark[1]
\affiliation{
  \institution{Wikimedia Foundation}
  \city{San Francisco}
  \country{USA}
}
\email{muniza.m.aslam@gmail.com}

\author{Ai-Jou Chou}
\affiliation{
  \institution{Wikimedia Foundation}
  \city{Berlin}
  \country{Germany}}
\email{achou@wikimedia.org}

\author{Miriam Redi}
\affiliation{
  \institution{Wikimedia Foundation}
  \city{London}
  \country{United Kingdom}}
\email{mredi@wikimedia.org}

\author{Diego Sáez Trumper}
\authornote{Corresponding author}
\affiliation{
  \institution{Wikimedia Foundation}
  \city{Barcelona}
  \country{Spain}}
\email{diego@wikimedia.org}

\renewcommand{\shortauthors}{Aitolkyn Baigutanova et al.}

\begin{abstract}
Wikipedia is a critical source of information for millions of users across the Web. It serves as a key resource for large language models, search engines, question-answering systems, and other Web-based applications. In Wikipedia, content needs to be verifiable, meaning that readers can check that claims are backed by references to reliable sources. This depends on manual verification by editors, an effective but labor-intensive process, especially given the high volume of daily edits. To address this challenge, we introduce a multilingual machine learning system to assist editors in identifying claims requiring citations. Our approach is tested in 10 language editions of Wikipedia, outperforming existing benchmarks for reference need assessment. We not only consider machine learning evaluation metrics but also system requirements, allowing us to explore the trade-offs between model accuracy and computational efficiency under real-world infrastructure constraints. We deploy our system in production and release data and code to support further research.
\end{abstract}




\maketitle

\section{Introduction}

Verifiability is one of Wikipedia’s core content policies~\cite{WikipediaCoreContentPolicies}, ensuring that information can be checked against reliable sources. In practice, this policy is frequently unmet, resulting in numerous claims without citations. As of September 2025, more than 550k articles on English Wikipedia contain one or more statements marked with a \emph{\texttt{\{\{Citation needed\}\}}} tag~\cite{WikiCitationNeedTemplate}. To address this issue, Wikipedia editors employ the \emph{\texttt{\{\{Citation needed\}\}}} tag~\cite{WikiCitationNeed}, which signals users that a statement requires a supporting source. Given the high volume of edits—approximately six per second as of 2024~\cite{WikipediaStats}—manually identifying and tagging uncited claims is prone to delays. Computational support is therefore essential to assist Wikipedia editors in maintaining verifiability standards. As pointed out by the Wikimedia Foundation's recently published AI Strategy~\cite{AIStrategy}, open
machine learning models that aid editors in preserving knowledge integrity are crucial to maintain Wikipedia’s quality.


Although claim detection on Wikipedia has been studied before, most prior research has focused on the English edition~\cite{10.1145/3308558.3313618,wright-augenstein-2020-claim}, with only a few case studies on smaller and less-resourced language editions. With Wikipedia spanning over 300 languages, there remains a need to tackle this problem in a multilingual context~\cite{johnson2022multilingual}. Furthermore, automated detection of unsupported claims on Wikipedia, which covers a broad range of topics, has not yet been explored in production-level settings. Given the need to adjust accuracy, latency, and resource availability, addressing the practical challenges of implementing such systems in real-world environments is crucial. 

\begin{figure*}[t]
  \centering
  \includegraphics[width=0.9\linewidth]{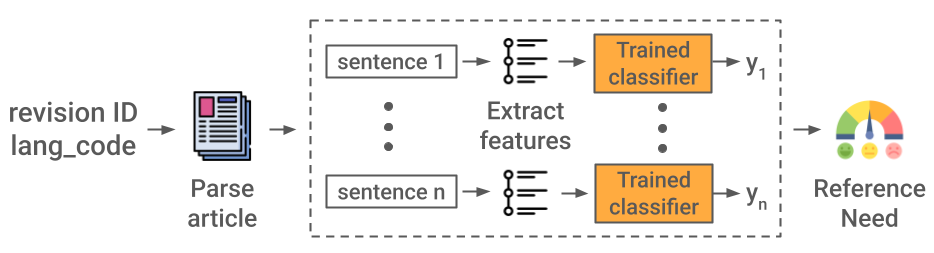}
  \caption{The Reference Need pipeline to compute the reference coverage score of an article revision.}
  \label{fig:intro}
\end{figure*}


In this work, we adopt the previously proposed reference need definition~\cite{baigutanova2023www} for a real-world product use case and introduce a novel machine learning system specifically designed to assess the reference coverage of Wikipedia articles at scale. The model is deployed with an initial use case aimed at assisting external re-users in evaluating the verifiability of Wikipedia articles. We further provide industrial insights into language modeling strategies by deploying the system under strict resource constraints, with particular focus on balancing model accuracy and system performance.


In summary, our contributions are as follows:
\begin{itemize}
    \item We introduce an open-source multilingual system for evaluating the reference need of Wikipedia articles, evaluated on the top 10 language editions by active user count. 
    \item We show the trade-offs between accuracy and latency when deploying language models in production under strict resource constraints. In this context, we explore model optimizations and compare small language models (SLMs) with large language models (LLMs) for this task, showing that SLMs may offer an optimal compromise between precision and serving time. 
    \item We release the training\footnote{\url{https://gitlab.wikimedia.org/repos/research/reference-quality/-/tree/research-notebooks/RN}} and inference code\footnote{\url{https://gitlab.wikimedia.org/repos/research/reference-quality/-/tree/classifiers/reference_quality/models/ref_need}}, along with the data\footnote{\url{https://analytics.wikimedia.org/published/datasets/one-off/reference-need}} used in our experiments, enabling further research in the field of information credibility on the Web.
\end{itemize}
\section{Related Work}
 Our system is designed to improve existing models in identifying reference needs on Wikipedia, leveraging recent advancements in language models.
 
\subsection{References on Wikipedia} The body of research work analyzing Wikipedia citation practices is growing,
including datasets of extracted citations~\cite{10.1162/qss_a_00226,kokash2024wikipedia} and studies of the source quality
~\cite{baigutanova2023cikm,yang2024polarization}. \citet{10.1145/3308558.3313618} contributed to predicting citation needs on Wikipedia by developing a taxonomy and a deep learning model, later integrated with Wikimedia community tools~\cite{chou2020citation}. The main limitation of these works is their focus on English, despite growing interest in multilingual research~\cite{johnson2022multilingual}. This is important for resource-limited language communities, where identifying citation needs is challenging due to scarce ground-truth data. 
\citet{HALITAJ2024100093}'s approach on three small Wikipedias has shown progress, but remains untested in production, leaving deployment challenges unaddressed.
 

\subsection{Masked Language Models} Language models have become pivotal in production-level natural language processing workflows~\cite{howard-ruder-2018-universal}. 
Masked language models (MLMs) like BERT~\cite{bert2019} and RoBERTa~\cite{DBLP:journals/corr/abs-1907-11692} have been successful in tasks such as sentiment classification~\cite{8947435}, readability assessment~\cite{trokhymovych-etal-2024-open}, and question answering~\cite{yang-etal-2019-end-end}. They scale well in multilingual contexts, efficiently handling over 100 languages with a single model~\cite{conneau-etal-2020-unsupervised}. However, MLMs require substantial resources for production-scale use. Techniques like knowledge distillation \cite{hinton2015distillingknowledgeneuralnetwork}, quantization~\cite{gong2014compressing}, and pruning~\cite{zhu2017prune} produced more efficient models. DistilBERT reduced model size by 40\% 
while retaining 97\% of language understanding,  
boosting speed by 60\%~\cite{sanh2020distilbertdistilledversionbert}. We explore the application of these methods to the reference need task. 

\subsection{Large Language Models} While modern LLMs can perform zero- and few-shot classification via simple prompting~\cite{wang2023largelanguagemodelszeroshot}, 
eliminating the need for extensive training, they require substantial computational resources, limiting their accessibility. Open-weight models, such as Llama~\cite{touvron2023llamaopenefficientfoundation,touvron2023llama2openfoundation} and Mistral~\cite{jiang2023mistral7b}, were released, but their deployment remains costly. Assessing LLM performance against smaller fine-tuned MLMs is essential to estimate their suitability for our task.

Moreover, recent research has shown the limitations of LLMs in applying complex Wikipedia policies, such as assessing content neutrality~\cite{ashkinaze2024seeing}, suggesting that this is a complex task that requires domain knowledge. Our work confirms these limitations and presents a comparison between large and small models, showing that smaller models can reach similar performance compared with their larger counterparts, require fewer resources, and offer faster serving time.
\section{Reference Need}
\citet{baigutanova2023www} defined Reference Need (RN) of a Wikipedia article as the proportion of uncited claims among those that require a reference. We adapt the original formulation, which considers all sentences in an article, to focus on a set of uncited sentences for prediction. Due to infrastructure constraints described below, we omit cited sentences from our evaluation pipeline because: 1) it notably reduces the number of sentences to evaluate, speeding up the inference process, and 2) evaluating sentences already reviewed by editors, who deemed them in need of a citation and provided one, is redundant. Since our model is trained to learn from that expertise, human-added citations are considered a gold standard. 

The reference need score is computed as follows:
\begin{equation}
RN = \frac{\sum_{i\in S_{\text{no-ref}}}y_i}{|S_{\text{ref}}| + \sum_{i\in S_{\text{no-ref}}}y_i}
\end{equation}
where $S_{\text{ref}}$ is the set of cited sentences in an article; $S_{\text{no-ref}}$ is the set of uncited sentences in an article; $y_i$ is the predicted label for uncited sentence $i$: $y=1$ if the sentence needs a citation and $y=0$ if the sentence does not need one.

Figure~\ref{fig:intro} shows the overall system that takes the revision ID of a Wikipedia article and the language code as input and outputs a reference need score between 0 and 1. A higher score indicates a higher proportion of sentences requiring citations. First, the revision content is retrieved through the MediaWiki API\footnote{\url{https://www.mediawiki.org/wiki/API}}, wikitext is parsed, and sentences are extracted. Next, target features used for classification are extracted from each uncited sentence and fed into the trained classifier, which predicts the probability that a sentence requires a citation. The predicted labels are combined to compute the overall reference need score of the revision.

\section{System Design and Deployment}
Our modeling approach relies on supervised learning of annotated data from Wikipedia articles. For our model to be deployed in production and classify articles at scale, multiple system requirements must be met, and masked language models are fine-tuned accordingly.

\subsection{System Requirements}
The infrastructure constraints and product needs of our system are as follows:
\begin{itemize}
    \item \textit{Serving time}: Wikipedia has around six edits per second. Therefore, fast models are needed to avoid creating long processing queues. Product and infrastructure standards aim for a maximum inference time of no more than 500 milliseconds.
    \item \textit{Inference infrastructure}: Due to challenges with adopting fully open-source GPUs, our model inference infrastructure relies only on CPUs. As a result, careful optimization is needed to achieve the required performance on this hardware.
    \item \textit{Multilingual support}: Our model aims to support at least the top 10 Wikipedia language editions by active editors. However, labeled data is scarce in some languages. To overcome this limitation, we rely on cross-lingual knowledge transfer techniques, using training data from the most widely used languages and testing on unseen languages. 
\end{itemize}

\subsection{Data Collection}
Determining whether a sentence needs a citation is complex, as it can be context-dependent and require domain knowledge. To address this challenge, we draw on the expertise of the Wikipedia community in curating knowledge. Wikipedia articles are manually ranked by quality, with the highest tier categorized as \textit{featured articles}\footnote{\url{https://en.wikipedia.org/
wiki/Wikipedia:Featured_articles}}, collectively decided by the editors. These are considered exemplary entries and serve as reference points for editors when writing or editing other articles. This strategy follows prior work that adopted a similar approach~\cite{10.1145/3308558.3313618}.

Using featured articles as our ground truth, we label sentences attributed to a source as needing a reference (positive class), and those without citation as not needing a reference (negative class). As the quality and quantity of featured articles may degrade with the size of a Wikipedia language edition, we limit the training data to five languages with a large amount of available data: English, French, German, Spanish, and Russian. We evaluate the final model on the top 10 languages by active users as of December 2024\footnote{\url{ https://en.wikipedia.org/
wiki/List_of_Wikipedias}}, additionally collecting data for Japanese, Persian, Italian, Portuguese, and Chinese editions. This set of languages covers the most active language communities while also representing a diverse range of language families and scripts, helping to understand how our model generalizes across languages. 

\begin{figure}[!t]
  \centering
  \includegraphics[width=\linewidth]{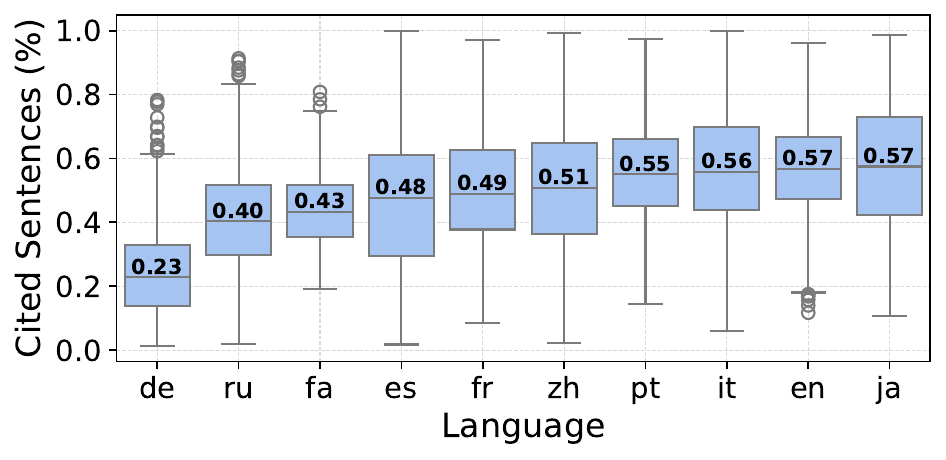}
  \caption{Proportion of sentences accompanied by a citation in featured articles by language.}
  \label{fig:data0}
\end{figure}

Table \ref{tbl:data0} outlines the dataset statistics for the 10 languages used in the evaluation. Featured articles, marked with a special wikitext template\footnote{\url{https://en.wikipedia.org/wiki/Wikipedia:Templates}} (e.g., \emph{\texttt{\{\{featured article\}\}}} in English Wikipedia), were extracted using regular expressions. The number of such articles ranges from 92 in Japanese to 6,508 in English. Figure~\ref{fig:data0} shows the proportion of cited sentences in featured articles, with more than five extracted sentences, across languages. Most have around half of the article sentences supported by citations, with a median proportion ranging from 0.40 to 0.57, while German has a lower proportion of 0.23. These cross-lingual differences align with findings observed in recent work~\cite{das2024language}. Manual inspection further reveals that articles in German often list literature at the end, not necessarily anchored to specific sentences~\cite{WikipediaLiteratur}.

\begin{table}[h!]
\caption{Statistics for the top 10 language editions by users.}
\label{tbl:data0}
\centering
\resizebox{0.9\columnwidth}{!}{%
\begin{tabular}{lcrrrr}
\toprule
\textbf{Language} & 
\makecell{\textbf{Used for} \\ \textbf{training}} & 
\makecell{\textbf{Active} \\ \textbf{users}} & 
\makecell{\textbf{Featured} \\ \textbf{articles}} & 
\makecell{\textbf{Cited} \\ \textbf{sentences}} & 
\makecell{\textbf{Uncited} \\ \textbf{sentences}} \\
\midrule
English (en) & Yes & 122,379  & 6,508  & 682,361  & 511,614  \\
French (fr)  & Yes & 17,728  & 2,145  & 327,240  & 353,463  \\
German (de)  & Yes & 17,426  & 2,848  & 199,538  & 685,580  \\
Spanish (es)  & Yes & 14,257  & 1,262  & 162,263  & 207,601  \\
Russian (ru)  & Yes & 9,225  & 1,902  & 271,976  & 418,045  \\
Japanese (ja)  & No & 12,427  & 92   & 22,005  & 14,462   \\
Portuguese (pt) & No & 8,423  & 1,439  & 178,856  & 148,806  \\
Chinese (zh)  & No & 7,066  & 923   & 69,736  & 78,684   \\
Italian (it)  & No & 7,494  & 555   & 66,113  & 54,879  \\
Persian (fa)  & No & 5,329  & 202   & 23,444  & 31,348  \\    
\bottomrule
\end{tabular}
}
\end{table}

\subsection{Data Processing}
We create our dataset using Wikimedia Data Lake’s \textit{Mediawiki wikitext current} table\footnote{\url{https://wikitech.wikimedia.org/wiki/Data_Platform/Data_Lake/Content}}, which is updated monthly and contains the article's latest revision ID, text, timestamp, page ID, and page title. The data used is from the May 2024 snapshot. The revision text is given in wikitext format, which we parse to plaintext using \textit{mwparserfromhell}\footnote{\url{https://github.com/earwig/mwparserfromhell}} and \textit{mwedittypes}\footnote{\url{https://github.com/geohci/edit-types}} libraries. 

Since references are key to our model, we modify the function converting wikicode to plaintext by keeping the location of reference tags, i.e., \textit{<ref>}, and citation templates, such as shortened footnotes and Harvard citations. Sentences containing these references are assigned a positive class, and the references are then removed for our final data. We also record the preceding and subsequent sentences in a given paragraph. When such sentences are missing, e.g., the target sentence is the first in a paragraph and lacks a preceding sentence, we record the preceding context as an empty string. The final dataset includes the following fields: language code, page ID, page title, revision ID, section title, target sentence, preceding and subsequent sentences, paragraph containing the target sentence, and citation label. 

We further eliminate duplicated sentences, which may arise when content is repeated or copied between related articles. Finally, we filter out sentences with fewer than six words and exclude sections that do not introduce new content (e.g., References, Further Reading, See Also)~\cite{10.1145/3308558.3313618}.
A full list of excluded sections by language is available in our code repository.

Field descriptions of the final dataset are given below:
\begin{itemize}
    \item \textit{wiki\_db}: the Wikipedia language edition 
    \item \textit{page\_id}: ID of a given article 
    \item \textit{page\_title}: title of a given article 
    \item \textit{revision\_id}: the ID of the corresponding revision, which is the latest edit at the time of dataset preparation
    \item \textit{section\_name}: name of the section where the target sentence is located
    \item \textit{sentence}: the target sentence
    \item \textit{next\_sent}: subsequent sentence of the target sentence within the paragraph
    \item \textit{prev\_sent}: preceding sentence of the target sentence within the paragraph
    \item \textit{paragraph}: the paragraph where the target sentence is located 
    \item \textit{label}: binary label representing whether the sentence is cited or not 

\end{itemize}

Figure~\ref{apd:fig0} shows a data sample of a target sentence that requires a citation (i.e., positive class).
\begin{figure}[h]
    \centering{
    \lstset{
        basicstyle=\ttfamily\footnotesize,
        frame=lines,
        breaklines=true,
        columns=fullflexible 
    }
    \begin{lstlisting}
{'wiki_db': 'enwiki',
 'page_id': 212989,
 'page_title': 'Western jackdaw',
 'revision_id': 1223095791,
 'section_name': 'systematics',
 'sentence': 'An archaic collective noun for a group of jackdaws is a "clattering"',
 'next_sent': 'Another name for a flock is a "train".',
 'prev_sent': '',
 'paragraph': 'An archaic collective noun for a group of jackdaws is a "clattering". Another name for a flock is a "train".',
 'label': 1 }
    \end{lstlisting}
    }
    \caption{Data sample of a sentence that needs a citation.}
    \label{apd:fig0}
\end{figure}

\subsection{Data Splitting} 
Data for training and testing are divided by page ID in each language edition to minimize article context sharing between data splits. The English, French, German, Spanish, and Russian editions are used for model training. We randomly sample 20K sentences balanced by the ground-truth label per language, resulting in a total of 100K sentences used for training, and similarly, 4K sentences per language, resulting in 20K sentences used for validation. The performance of the resulting classifier is evaluated on a total of 30K sentences from a holdout set of articles in the 10 languages listed in Table~\ref{tbl:data0}.

\subsection{Model Training}
We fine-tuned the distilled version of the multilingual BERT model~\cite{bert2019} for the citation needed classification task. 
The input for the model includes the language code, the section name where a sentence is located, the sentence itself,  the preceding and subsequent sentences separated by a separator token, and the label that describes whether the sentence in our training data had a reference or not. We use the transformers package~\cite{wolf2020transformers} for fine-tuning from the pre-trained model. The final model uses \textit{distillbert-base-multilingual-cased} as a base model with a maximum input sequence length of 128 tokens (see Section~\ref{sec:experiments}). Following recommendations on optimal hyperparameters~\cite{bert2019}, we set the learning rate to 1e-5, a weight decay to 0.01, used a batch size of 16, and performed training for 3 epochs. Training resources are limited to an AMD Radeon Pro WX 9100 16GB GPU.

\subsection{Model Deployment}
The Reference Need system is deployed using KServe\footnote{\url{https://wikitech.wikimedia.org/wiki/Machine_Learning/LiftWing/KServe}}, a standard model inference platform on Kubernetes\footnote{\url{https://kubernetes.io}}, ensuring scalability. 
The model server runs as an asynchronous, containerized microservice within a dedicated ML cluster. The service is configured with resources of 4 CPUs and 6 GB of memory and is accessible through a public API gateway\footnote{\url{https://api.wikimedia.org/wiki/Lift_Wing_API/Reference/Get_reference_need_prediction}} (see Appendix~\ref{appendix:api}). 

\section{Experiments}
\label{sec:experiments}

Our system is a supervised classification model that predicts whether a sentence needs to be supported by a citation. We test different input feature sets and language model configurations, evaluating performance based on model accuracy and efficiency. We use AUC-ROC as a primary metric in the following experiments, as it is threshold-independent. Furthermore, as the final model runs in a CPU-only production environment, computational performance results are based on CPU inference, with the number of cores set to four to match deployment conditions. For LLM comparison, we report latency using one additional GPU (AMD Instinct MI 210) due to the model sizes.

\subsection{Input Selection}
\label{subsec:input-res}
First, we run experiments to compare different inputs to the model. In particular, we explore how the following features affect model capabilities: language, the title of the section where the sentence is located, the paragraph, and the surrounding sentences. 

We fix the pre-trained model to the distilled version of mBERT with a context size of 128 for this comparison. We begin by providing the model with only the target sentence (S). Subsequently, we incorporate additional contextual information, including the language code (L), the section title where the sentence appears (ST), the sentence triplet comprising the target sentence, the following sentence, and the preceding sentence (SNP), as well as the entire paragraph containing the target sentence (para). As shown in Figure~\ref{fig:res0}, the L+ST+SNP combination (i.e., language code, section name, and sentence triplet) outperforms other options, specifically improving from a single sentence input by 0.111 in ROC-AUC. As a result, we fix the input for subsequent experiments to this combination, with features separated by the \textit{[SEP]} token.

\subsection{Model Selection}

Next, we evaluate the performance of fine-tuned MLMs for the reference need detection task at the sentence level. Given the multilingual context of the task, the base models tested include \textit{bert-base-multilingual-cased}, \textit{distilbert-base-multilingual-cased}, and \textit{xlm-roberta-base}. BERT models were introduced by Google researchers in 2019 and were mainly trained on Wikipedia data and have been largely used as primary SLMs~\cite{bert2019}. \textit{distilbert-base-multilingual-cased} is a distilled version of the original BERT, aiming to make it smaller and faster. RoBERTa models were developed by Meta and were trained on CommonCrawl\footnote{\url{https://commoncrawl.org}} data~\cite{DBLP:journals/corr/abs-1907-11692} and are commonly used as an alternative to BERT.

\begin{figure*}[!t]
\centering
    \begin{subfigure}[b]{0.45\linewidth}\centering
    \includegraphics[width=\linewidth]{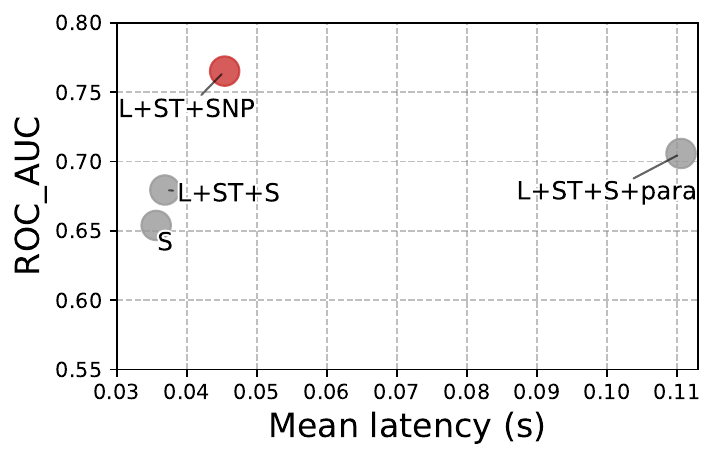} 
    \caption{By model input: language (L), section title (ST), target sentence (S), paragraph (para), and sentence - next sentence - previous sentence triplet (SNP).}
    \label{fig:res0}
    \end{subfigure}
  \hfill
  \begin{subfigure}[b]{0.45\linewidth}\centering
  \includegraphics[width=\linewidth]{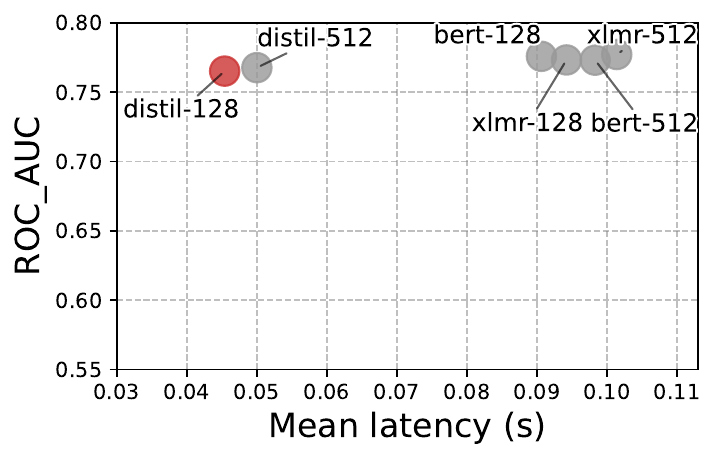}
  \caption{By base model and input sequence length: DistilBERT multilingual (distil), BERT multilingual (bert), and XLM-Roberta (xlmr). Each with 128 and 512 input tokens.}
  \label{fig:res1}
  \end{subfigure}

\begin{subfigure}[b]{0.45\linewidth}\centering
    \includegraphics[width=\linewidth]{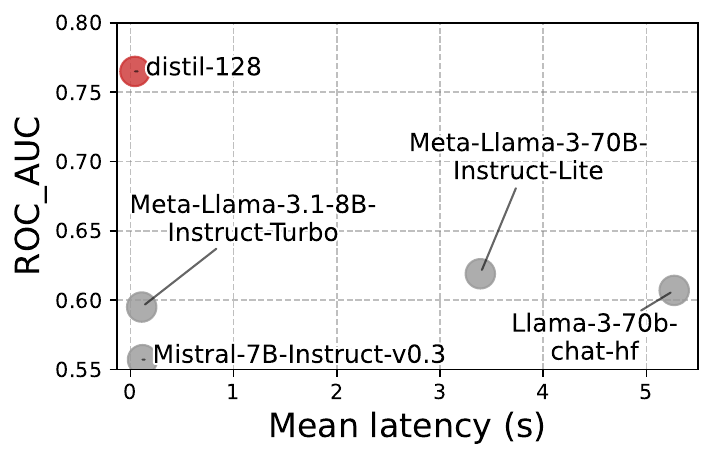}
    \caption{Zero-shot LLMs.}
    \label{fig:res4}
    \end{subfigure}
\hfill
  \begin{subfigure}[b]{0.45\linewidth}\centering
  \includegraphics[width= \linewidth]{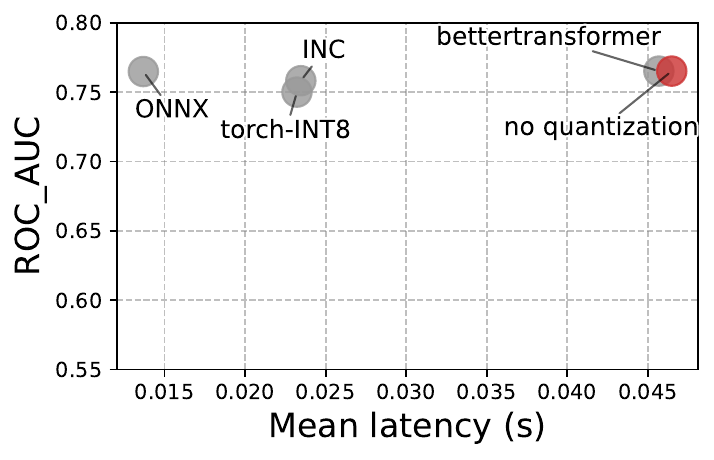}
  
  \caption{By optimization method.}
  \label{fig:res3}
    \end{subfigure}
\caption{Performance and latency comparison. The red dot in all subfigures represents the fine-tuned distil-128 model.}
\end{figure*}

Figure~\ref{fig:res1} compares base language models with input sequence lengths of 128 and 512 (e.g., distil-128 and distil-512) in terms of accuracy and time taken per prediction. In Table \ref{tbl:res0}, we report AUC-ROC, accuracy, F1-score, precision, and recall for performance measurements along with the inference time per sentence. We additionally add 95\% confidence intervals (CI) for the AUC-ROC metric, which are defined as two standard deviations around the resulting bootstrap estimate~\cite{efron1994introduction}.

We observe that larger models and longer input sequence lengths allow for better accuracy. On the other hand, the distilled version with a smaller input length performs on par, but the latency of the model reduces around twofold. Therefore, we select the fine-tuned distilled BERT multilingual model with an input sequence length of 128 tokens as a base model for our final classifier. 

\begin{table*}[!t]
\caption{Reference Need model performance sorted by AUC-ROC (the final model is underlined and the best performance values are bolded).
}
\label{tbl:res0}
{
\centering
\resizebox{\textwidth}{!}{%
\begin{threeparttable}
\begin{tabular}{ccccccccc} 
\toprule
\textbf{Base model}          & \textbf{LM family} &  \textbf{Quantize}  & \textbf{AUC-ROC} & \textbf{Acc} & \textbf{F1-score} & \textbf{Precision} & \textbf{Recall} &  \textbf{\begin{tabular}[c]{@{}c@{}}Mean\\ latency (s)\end{tabular}} \\ \midrule
Mistral-7B-Instruct-v0.3 & LLM & FP16                   & 0.557 $\pm$ 0.007           & 0.521          & 0.649             & 0.512              & \textbf{0.886}           &     0.118                                                               \\ 
Meta-Llama-3.1-8B-Instruct-Turbo & LLM & FP8                 & 0.595 $\pm$ 0.006            & 0.565          & 0.564             & 0.565              & 0.564           &     0.110                                                               \\ 
Llama-3-70b-chat-hf & LLM & FP16                & 0.607 $\pm$ 0.006            & 0.560          & 0.642             & 0.541              & 0.790           &              5.268                                                      \\ 
Meta-Llama-3-70B-Instruct-Lite &  LLM& INT4                & 0.619 $\pm$ 0.006           & 0.566          & 0.643             & 0.546              & 0.783           &        3.390                                                            \\ 
Citation-Needed~\cite{10.1145/3308558.3313618}\tnote{*} & SLM & - & 0.650 & 0.590        & 0.491             & 0.647              & 0.395                    & -                                                              \\ 
\underline{distilbert-128} &SLM & \underline{torch-INT8} & \underline{0.750 $\pm$ 0.006}   & \underline{0.676} & \underline{0.696}    & \underline{0.655}     & \underline{0.743}  & \underline{0.023}                                                     \\ 
distilbert-128          &SLM & Intel-NC            & 0.758 $\pm$ 0.006            & 0.687          & 0.689             & 0.684              & 0.694           & 0.023                                                              \\ 
distilbert-128          &SLM & ONNX                & 0.765 $\pm$ 0.005            & 0.688          & 0.693             & 0.686              & 0.701           & \textbf{0.014}                                                              \\ 
distilbert-128         & SLM& bettertransformers  & 0.765 $\pm$ 0.005            & 0.692          & 0.692             & 0.693              & 0.691           & 0.046                                                              \\ 
distilbert-128          & SLM& -                   & 0.765  $\pm$ 0.005            & 0.692          & 0.692             & 0.693              & 0.691           & 0.045                                                              \\ 
distilbert-512          & SLM& -                   & 0.768 $\pm$ 0.005            & 0.694          & 0.688             & 0.702              & 0.676           & 0.050                                                              \\ 
bert-512               & SLM& -                   & 0.773 $\pm$ 0.005            & 0.701          & \textbf{0.704}             & 0.698              & 0.710           & 0.098                                                              \\ 
xlm-roberta-128 &SLM & -                   & 0.773 $\pm$ 0.005            & 0.701          & 0.702             & 0.699              & 0.706           & 0.094                                                              \\ 
bert-128                &SLM & -                   & 0.776 $\pm$ 0.005            & 0.702          & 0.697             & 0.708              & 0.687           & 0.091                                                              \\ 
xlm-roberta-512  & SLM & -                   & \textbf{0.777} $\pm$ 0.005            & \textbf{0.707}          & 0.700             & \textbf{0.718}              & 0.682           & 0.101                                                              \\ \bottomrule
\end{tabular}

\begin{tablenotes}
    \item[*] Citation-Needed was evaluated only on the subset of English articles due to the model's limitations.
\end{tablenotes}
    \end{threeparttable}
}}
\end{table*}

Additionally, we compare our results to the existing baseline, Citation-Needed model~\cite{10.1145/3308558.3313618}. Since it supports only English Wikipedia, we evaluated it on English sentences from our test data, excluding other languages. The results confirm that our method outperforms Citation-Needed significantly.

\begin{figure}[t]
  \centering
  \includegraphics[width=\linewidth]{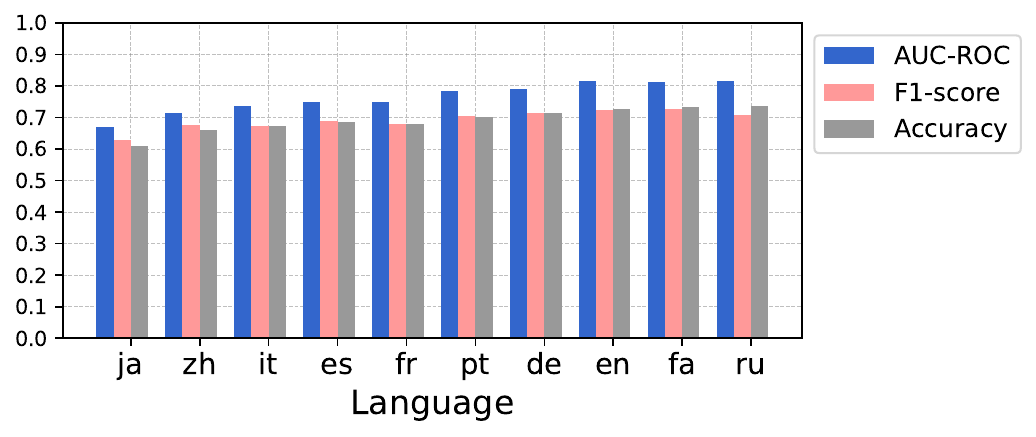}
  \caption{Performance by language using distilbert-128.}
  \label{fig:res2}
\end{figure}

While our Reference Need model is trained on five languages, its performance is evaluated on 10 languages, including previously unseen ones. We present model performance across languages in Figure \ref{fig:res2}. Notably, the model performs on par in unseen languages, particularly Portuguese, Persian, and Italian, while Chinese and Japanese show relatively lower performance. Performance breakdowns by language are available in Appendix \ref{apd:lang-result}.


\subsection{Comparison with LLMs}
We further explore a zero-shot approach using LLMs. This approach offers significant advantages, as it does not require additional training data, it is user-friendly, and can be applied across multiple languages. While LLMs' high inference computational load currently exceeds our infrastructure constraints, we still evaluate their performance for our task to compare their effectiveness with our proposed system.

We select those LLMs that could realistically eventually run in a resource-constrained setting such as ours. We utilize the external platform Together.AI\footnote{\url{https://www.together.ai}} to access LLMs for inference and experimented with the following models: \textit{Llama-3-70b-chat-hf}, \textit{Llama-3-70B-Instruct-Lite}, \textit{Llama-3.1-8B-Instruct-Turbo}, and \textit{Mistral-7B-Instruct-v0.3}. \textit{Turbo} indicates FP8 quantization, \textit{Lite} refers to INT4, and \textit{hf} denotes the original weights. 

Our goal is not only to obtain a binary prediction but to assess the confidence level of the LLMs' outputs. We achieve this with a manual verbalizer approach~\cite{schick-schutze-2021-exploiting} and an instruction-based prompt to evaluate if the text requires a reference, explicitly asking the model to return \textit{"yes"} or \textit{"no"}. This evaluation is based on the target sentence, the following sentence, the preceding sentence, the corresponding language code, and the section name, following the selected input configuration. We run inference twice, each time placing a different variant of the answer directly in the prompt (see Appendix~\ref{appendix:prompt} for prompt details). The final classification label and probability score are then determined by applying the softmax activation function to the log probabilities of the \textit{"yes"} and \textit{"no"} answer tokens from each run.

The final results are shown in Figure~\ref{fig:res4}. We observe that the fine-tuned DistilBERT model configuration significantly outperforms LLM-based configurations. Among LLMs, smaller models (7B parameters) perform slightly worse than larger ones (70B parameters), and quantization does not considerably impact the results. Given that the inference latency of LLM-based approaches is substantially higher than that of SLMs, more advanced prompting strategies, such as few-shot and Chain-of-Thought, would be infeasible under our constrained infrastructure. Moreover, prior work by \citet{ashkinaze2024seeing} shows that few-shot approaches provide limited gains when applying LLMs to Wikipedia policy-related tasks. 

Overall, these findings suggest that fine-tuning small language models might still be more effective than using pre-trained LLMs out of the box for complex tasks like citation need prediction. Nevertheless, even without task-specific tuning, LLMs outperform random predictions, indicating that their performance could be improved with additional tuning in future work. 

\subsection{Model Optimization}

We finally improve inference speed using post-training optimizations like quantization and pruning. Quantization reduces the bit width of model parameters and activations by applying a scaling factor, lowering memory usage and accelerating computations. We use dynamic quantization, where weights are pre-quantized, while activations are quantized during inference. Pruning removes redundant nodes to further reduce model size. We test whether the accuracy of our model has been maintained with optimizations. 

We evaluate several methods, including PyTorch quantization\footnote{\url{https://docs.pytorch.org/docs/stable/quantization}}, exporting to ONNX\footnote{\url{https://onnxruntime.ai/}} and enabling graph optimizations before applying quantization, implementing quantization using Intel Neural Compressor (NC)\footnote{\url{https://github.com/intel/neural-compressor}}, and better transformer optimization\footnote{\url{https://huggingface.co/docs/optimum/en/bettertransformer/overview}}.

Figure \ref{fig:res3} shows that the optimized versions maintain accuracy comparable to the original model while reducing latency by more than twice. Our production environment is currently set up for PyTorch models, so despite the ONNX model achieving the lowest latency, we chose the PyTorch quantized model for deployment. Running ONNX would require integrating onnxruntime and modifying the inference code to support ONNX-specific abstractions, which is left for future work.

\section{Conclusion}

In this paper, we introduced an open-source multilingual system for assessing reference need in Wikipedia articles.
By analyzing the trade-offs between model accuracy and inference latency in production settings, we demonstrated the practicality of SLMs in balancing performance and resource constraints. 
We further examined optimization techniques and compared our final model with LLMs, showing that zero-shot LLM performance may be lacking compared to fine-tuned SLM alternatives for citation need detection task. 

While latency measurements in this work are reported under CPU-only settings, performance could be further improved with GPU resources. As the system is already deployed in production, future work includes qualitative and quantitative evaluation with Wikipedia editors to better understand its perceived utility and real-world usage. Ultimately, we aim for this system to support editors in improving the verifiability of content across multiple language editions of Wikipedia.

\section{Limitations}

In this work, we have focused on well-established SLMs like DistilBERT, which have proven robust in production but have not been updated in recent years. While newer models like ModernBERT have been released~\cite{modernbert}, they are currently lacking the broad multilingual support required for our use case. 

Our approach also faces computational constraints. Processing time scales linearly with the number of uncited sentences, making longer articles more time-consuming to assess. To address this limitation, future work could explore chunk-based strategies for inference to reduce latency and improve scalability compared to per-sentence processing.

Finally, although LLMs might achieve better performance with extensive prompt optimization, they remain unsuitable for our current technical needs. Our LLM choices were limited by our infrastructure constraints. This work has focused on evaluating systems that could be applied in real-world scenarios within the context predetermined by our technical framework at the time of releasing this system. Nevertheless, the rapid evolution of LLMs requires us to keep testing new models and evaluating novel opportunities they might bring.

\section{Ethical Considerations}
This work has several caveats. The proposed model relies on the distilled version of multilingual BERT, which might contain inherent biases~\cite{jentzsch2022gender}. 
Moreover, while the model was designed to work in around 100 languages, its performance has been tested only in the top 10 language editions. These editions include languages from different alphabets and regions, each with well-maintained featured articles. However, the classifier should be used with precautions in other Wikipedia editions, as the quality of predictions might decrease for underrepresented languages. Future work on low-resource languages is needed to help advance knowledge equity. Additionally, the cross-lingual learning transfer approach was mainly based on English and other Western languages, which may have different citation styles and policies compared to other languages. This might be considered a type of cultural imposition over smaller Wikipedia language communities. 

Lastly, misuse of this tool might involve bad-faith editors hiding their intentions of introducing low-quality or fake content, making it harder for patrollers and automated tools to detect such behavior. Moreover, given the unique nature of Wikipedia, this tool was specifically designed for use within the platform and may not generalize well to other domains. To clarify the intended use cases and mitigate misuse, we will release a model card with detailed documentation~\cite{modelcard2019}.

\bibliographystyle{ACM-Reference-Format}
\balance
\bibliography{reference}

\appendix
\section{Prompting details}
\label{appendix:prompt}
Here, we present additional technical details to help interpret the results of LLMs evaluation. Figure~\ref{prm:verbalizer} provides the prompt template used for zero-shot reference need classification task with a manual verbalizer.

\begin{figure}[h]
    \centering
    \lstset{
        basicstyle=\ttfamily\footnotesize,
        frame=single,
        breaklines=true,
        columns=fullflexible 
    }
    \begin{lstlisting}
    
    You are an experienced Wikipedia editor.

You are provided with a piece of text from a Wikipedia article enclosed in `<<>>`.
The language code of the article is `{{language}}`, and the section name is `{{section}}`.
Optionally, you may also be provided with context before 
the text snippet (enclosed in `< context before >`) and 
after the text snippet (enclosed in `<<< context after >>>`).

Follow these guidelines while making your assessment:

### Text that REQUIRES a citation:
- Text that includes facts likely to be challenged or not considered common knowledge.
- Statements that present opinions, analyses, or claims that need verification.
- Statistics, data, or direct quotations.
- Assertions that could be considered original research.

### Text that DOES NOT REQUIRE a citation:
- Text that is widely accepted as common knowledge.
- Content within sections where the main topic is already properly referenced.
- Plot summaries for works of fiction.
- Statements that are already supported by a preceding or nearby citation.
- Other cases where a citation is clearly redundant or unnecessary.

**Your task is to determine whether the text snippet in `<<>>` 
requires a citation according to Wikipedia's citation policies.**

ANSWER FORMAT:
yes - Text to assess snippet requires a citation.
no - Text to assess snippet does NOT require a citation.

INPUT: 
{{context_b}}
Text to assess: <<{{text}}>>
{{context_a}}
Answer: "yes" {or "no"}

    \end{lstlisting}
    \caption{A prompt template is used for reference-needed classification with a manual verbalizer. The last line includes either \textit{"yes"} or \textit{"no"} to extract the log probability for each token representing the corresponding label.}
    \label{prm:verbalizer}
\end{figure}

\section{Performance by Language}
\label{apd:lang-result}
\begin{figure}[h!]
  \centering
  \includegraphics[width=\linewidth]{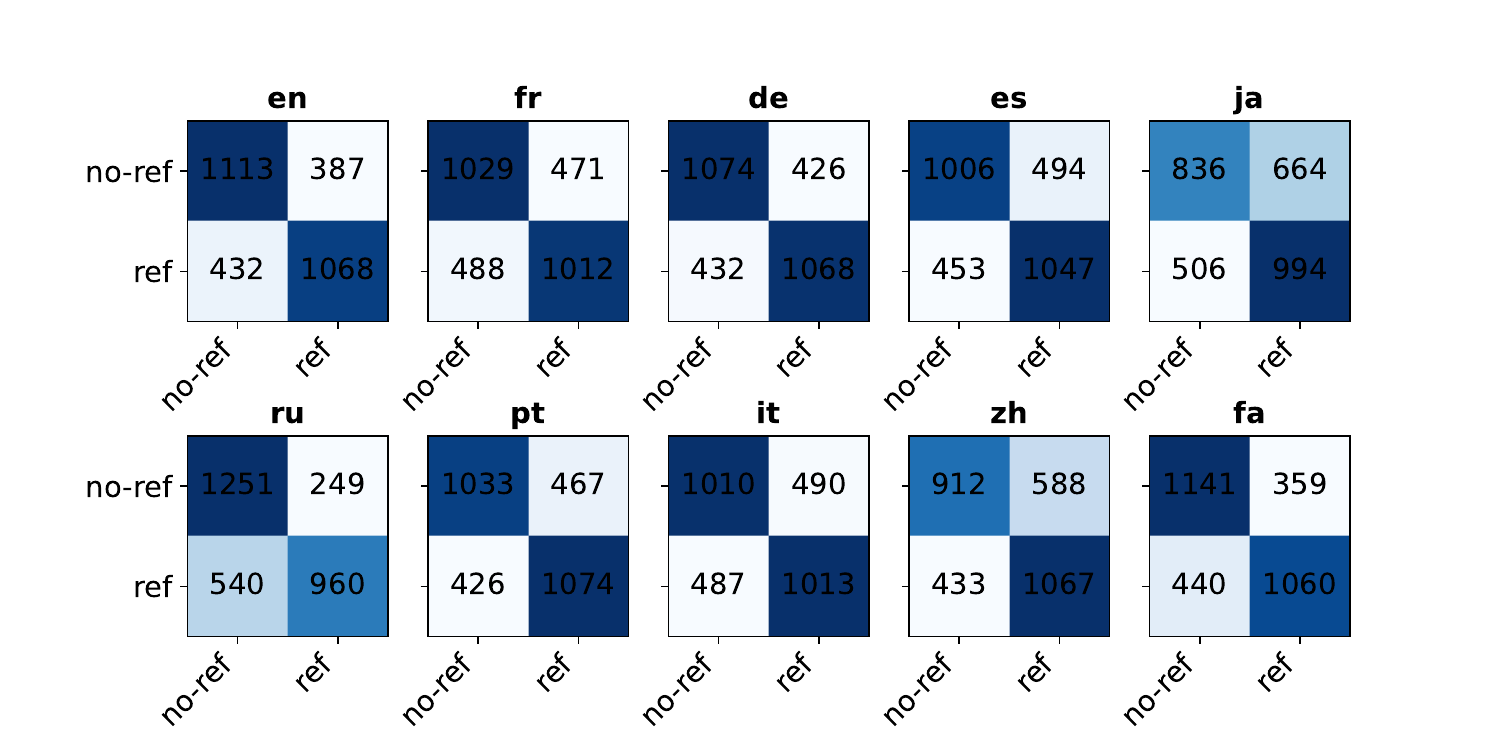}
  \caption{Confusion matrices by language using distil-128 model without optimizations.}
  \label{fig:app1}
\end{figure}

Figure~\ref{fig:app1} shows the confusion matrices for each of the 10 languages used in the evaluation.

\section{Reference Need service}
\label{appendix:api}
Here, we show an example output of an API call to the Reference Need service. 
Given the input \texttt{\{"rev\_id": 1242378206, "lang": "en"\}}, the output is shown in Figure~\ref{apd:api_out}.

\begin{figure}[h!]
    \centering{
    \lstset{
        basicstyle=\ttfamily\footnotesize,
        frame=lines,
        breaklines=true,
        columns=fullflexible
    }
    \begin{lstlisting}
{"model_name":"reference-need",
"model_version":0,
"wiki_db":"enwiki",
"revision_id":1242378206,
"reference_need_score":0.16666666666666666}   
    \end{lstlisting}
    }
    \caption{Output from the API call to the Reference Need model.}
    \label{apd:api_out}
\end{figure}

\end{document}